\documentclass{osa-article}

\journal{oe}

\articletype{Research Article}

\begin{document}

\title{Evidence of nonlinear diffusion in KTP waveguides}

\author{Laura Padberg,\authormark{1,*} Matteo Santandrea,\authormark{1}, Michael R\"using,\authormark{2} Julian Brockmeier,\authormark{3} Peter Mackwitz,\authormark{3} Gerhard Berth,\authormark{3} Artur Zrenner,\authormark{3} Christof Eigner,\authormark{1} and Christine Silberhorn\authormark{1}}

\address{\authormark{1}Integrated Quantum Optics, Paderborn University, Warburger Strasse 100, 33098 Paderborn, Germany\\
		\authormark{2}Institute of Applied Physics, Technical University Dresden, 01062 Dresden, Germany\\
		\authormark{3}Nanostructure Optoelectronics, Paderborn University, Warburger Strasse 100, 33098 Paderborn, Germany}

\email{\authormark{*}laura.padberg@uni-paderborn.de} 



\begin{abstract}
Integrated $\chi^{(2)}$ devices are a widespread tool for the generation and manipulation of light fields, since they exhibit high efficiency, small footprint and the ability to interface them with fibre networks. Surprisingly many things are not fully understood until now, in particular the fabrication of structures in potassium titanyl phosphate (KTP). A thorough understanding of the fabrication process and analysis of spatial properties is crucial for the realization and the engineering of high efficiency devices for quantum applications. 
In this paper we present our studies on rubidium-exchanged waveguides fabricated in KTP. Employing energy dispersive X-ray spectroscopy (EDX), we analysed a set of waveguides fabricated with different production parameters in terms of time and temperature. We find that the waveguide depth is dependent on their widths by reconstructing the waveguide depth profiles. Narrower waveguides are deeper, contrary to the theoretical model usually employed.
Moreover, we found that the variation of the penetration depth with the waveguide width is stronger at higher temperatures and times. We attribute this behaviour to stress-induced variation in the diffusion process. 
\end{abstract}

\section{Introduction}
Nonlinear waveguide structures are essential for state generation and manipulation and provide a compact solution for interfacing components of a quantum network operating at different wavelengths \cite{maring,rutz}.
Moreover, waveguides offer stronger field confinement, thus achieving much higher efficiencies per unit length then their bulk counterparts. 
Among the different technological platforms like the well-known lithium niobate, KTP is especially interesting because it possesses unique dispersion properties which make it highly interesting for quantum state generation \cite{harder}. Nevertheless the behavior of KTP during waveguide fabrication and periodic poling is not fully understood.
KTP offers large nonlinear optical and electro-optical coefficients \cite{bierlein3}, a wide transparency range extending well into the UV region \cite{bierlein3}, high damage thresholds and low photorefraction \cite{chu}. Moreover, its anisotropy of the domain growth facilitates the poling of the material \cite{canalias}, thus enabling the realization of sub-micron structures.
Finally, the possibility to fabricate waveguides \cite{bierlein2} makes this platform ideal for the development of integrated devices.
For these reasons, the development of homogeneous waveguides in potassium titanyl phosphate is of great interest. However, waveguide structures are much more sensitive to inhomogeneities and fabrication imperfections, that can result in increased losses and reduced efficiency. \\
Several techniques have been used to produce waveguides in KTP, e.g. ion-exchange \cite{bierlein2}, ridge waveguides \cite{volk,eigner} or laser writing \cite{laurell}.
Ridge waveguides offer a high mode confinement due to the substrate air boundaries, but the losses are rather high. They were measured to be not lower than 1.3dB/cm \cite{volk}. The most promising up-to-date technology makes use of rubidium exchanged channel waveguides, as they have shown remarkably low losses of 0.67 dB/cm \cite{ansari}. The fabrication is known to be non trivial \cite{bierlein2}, but no rigorous investigation of fabrication dynamics have been conducted so far. Using energy dispersive X-ray spectroscopy (EDX) we have analysed the concentration profiles of planar and channel Rb:KTP waveguides and characterized the diffusion profile of rubidium for different fabrication conditions. The EDX measurements have revealed an unexpected dependence of the diffusion depth to the waveguide width.
With the help of Raman analysis, we show that this unusual dynamics can be related to stress present in the waveguide. 

\section{Waveguide fabrication and EDX methodology}
\begin{figure}[ht!]
		\centering
		\includegraphics[width=12.5cm]{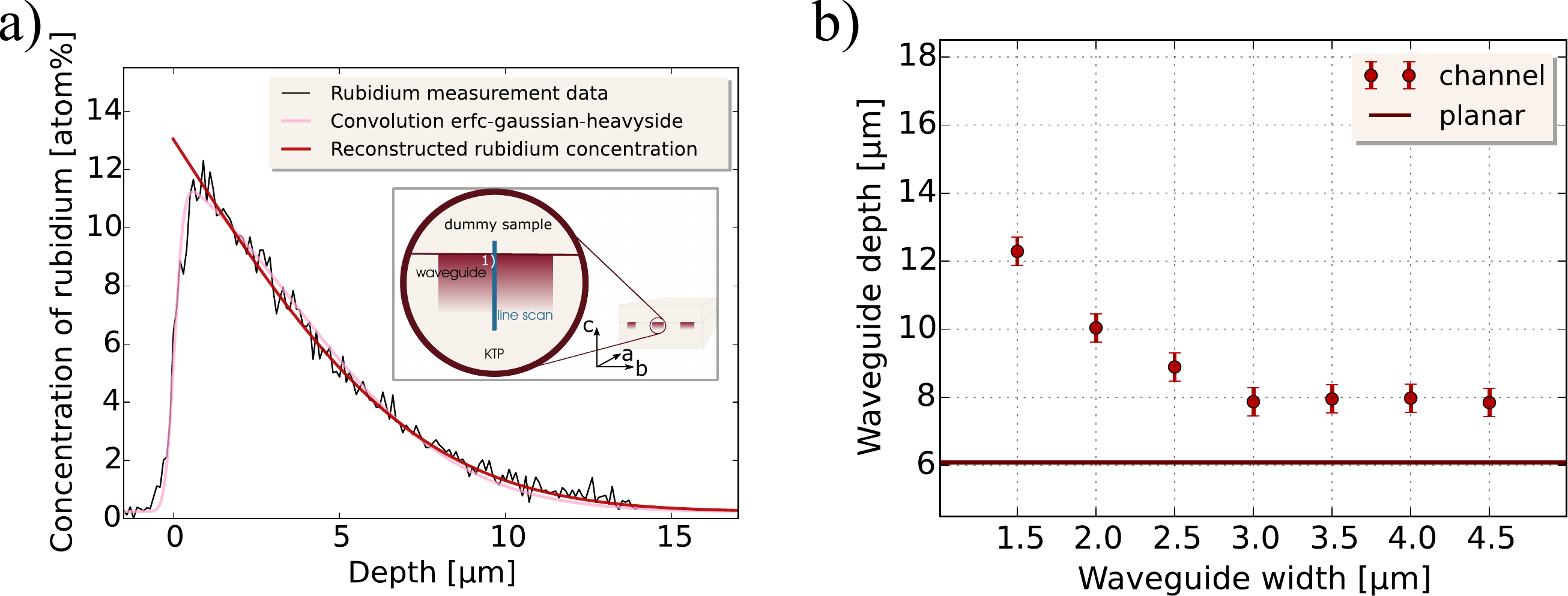}
		\caption{a) Measured rubidium concentration in black and fit of the convolution of the expected complementary error, gaussian and Heaviside function in rose. The red line shows the reconstructed concentration profile and the calculated depth for a 3.0\, \textmu m channel waveguide. The inset shows how the waveguides have been measured. With a linescan at the b-c plane the concentration profile in depth has been reconstructed. b) Waveguide depth depending on the waveguide width of a sample produced at 330$^\circ$C for 60\,min. The dark red line shows the depth of planar waveguides produced at 330$^\circ$C for 60\,min.}
		\label{fig:edx}
	\end{figure}

For the waveguide profile analysis we cut a commercially available wafer into smaller pieces of (10x6x1)mm along the crystallographic directions a-b-c. We fabricated waveguides in KTP by a rubidium-potassium exchange, which results in a local in-diffusion of rubidium in the KTP bulk material. 
The rubidium increases the refractive index of the material which allows waveguiding in the in-diffused region \cite{bierlein2}.
The Rb-K exchange is performed by immersing the sample in a rubidium nitrate (RbNO$_3$), potassium nitrate (KNO$_3$) and barium nitrate (Ba(NO$_3$)$_2$) melt for a specific time and at a specific temperature. 
The simplest description of rubidium in-diffusion consists in modelling the melt as an infinite reservoir for the diffusion process and assuming a concentration-independent diffusion coefficient $D$. The resulting concentration profile of the exchanged Rb ions follows a complementary error function \cite{crank}
	\begin{align}
		c(z,t)=c_0 \cdot \textrm{erfc}\left(\frac{z}{2 \cdot [D(T) \cdot t]^{1/2}}\right) \, . \label{eq:c}
	\end{align}
where $c$ is the concentration of rubidium ions inside the crystal, $c_0$ is the surface concentration of rubidium and $t$ is the exchange time. The point $z=0$ describes the surface of the crystal, while $z>0$ describes the distance into the sample and $c$ describes the concentration. \\
Planar waveguides are produced by directly immersing the sample into the melt. To fabricate channel waveguides, a titanium mask  is photolithographically patterned on the KTP surface perpendicular to the c-axis. In the investigated samples, the channel width was varied from 1.5\,\textmu m to 4.5\,\textmu m on the -c face. \\ 
We determined the concentration profiles of the waveguides using EDX. We scanned a line in the cross section of the guiding waveguides named the b-c plane, see inset Fig. \ref{fig:edx}a). In this way, we were able to retrieve the Rb concentration profile at the centre of the waveguide. The measured profile shows a trend compatible with the 1D diffusion model described above. The measured EDX data are a convolution of different process parameters such as the geometry of the excitation beam and effects due to the edge of the sample. For the determination of the concentration profile, we convolute the expected complementary error function with a Gaussian and a Heaviside function and fit this to the measured data, see rose line in Fig. \ref{fig:edx} a). The Heaviside function describes the edge of the sample. For the spread of the electron beam which has a Gaussian distribution, we exploit a gaussian function which describes a 220\,nm wide excitation beam. The dark red line shows the deconvoluted complementary error function and the reconstructed rubidium concentration profile of a waveguide.  
From this reconstructed concentration profile we calculate the penetration depth that can be defined as $c(z,t)/c_0 = \textrm{erfc}(1)$, that is \cite{korkishko} 
	\begin{align}
		d = 2 \cdot [D(T) \cdot t]^{1/2} \label{eq:d} \, .
	\end{align}
	
\section{Waveguide analysis} \label{analysis}
\begin{figure}[ht!]
		\centering
		\includegraphics[width=13.5cm]{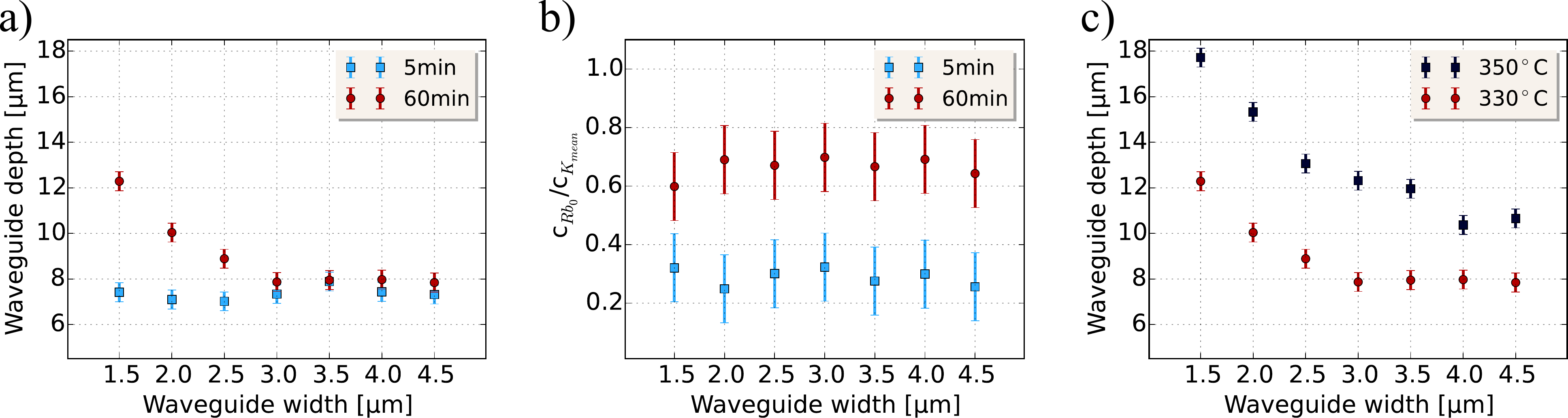}
		\caption{a) Waveguide depth depending on the waveguide width of a sample produced at 330$^\circ$C for 5\,min (light blue) and 60\,min (red) in 97\% mol $\mathrm{RbNO_3}$, 1\% mol $\mathrm{Ba(NO_3)_2}$, 2\% mol $\mathrm{KNO_3}$. The uncertainty of the waveguide depth is calculated via standard deviation to $\pm$ 0.4\,\textmu m. b) Ratio of exchanged potassium ions with rubidium ions at the surface depending on the waveguide width of a sample produced at 330$^\circ$C for 5\,min (light blue) and for 60\,min (red) in 97\% mol $\mathrm{RbNO_3}$, 1\% mol $\mathrm{Ba(NO_3)_2}$, 2\% mol $\mathrm{KNO_3}$. The uncertainty of the ratio is calculated to $\pm 0.12$.  c) Waveguide depth depending on the waveguide width of a sample produced at 330$^\circ$C for 60\,min (red) and 350$^\circ$C for 60\,min (dark blue) in 97\% mol $\mathrm{RbNO_3}$, 1\% mol $\mathrm{Ba(NO_3)_2}$, 2\% mol $\mathrm{KNO_3}$.}
		\label{fig:330_5_800}
	\end{figure}

We studied the diffusion properties and dynamics of Rb in KTP. For that we investigated the penetration depth of planar and channel waveguides produced at 330$^\circ$C for 60\,min. Using these parameters it is possible to fabricate waveguides, which guide only one mode at infrared wavelengths \cite{ansari}.
\paragraph{Energy dispersive X-ray spectroscopy} $~~$\\
The results of the EDX measurements of planar and channel waveguides with different width are displayed in Fig. \ref{fig:edx}b). Both waveguide types are fabricated with the same parameters. If we compare the penetration depths of channel waveguides and planar waveguides, we see that the channel waveguides are at least 2\,\textmu m deeper even though both of them are produced with the same fabrication parameters. 
Moreover, we can clearly see a trend towards increasing waveguide depth with decreasing width of the waveguides. 
These observations clearly demonstrate that the simple model is not capable of describing the whole diffusion dynamics of Rb in KTP. The expected erfc function in Fig. \ref{fig:edx}a) valids to assume a 1D model, but the model is not valid in comparison for different geometries in Fig. \ref{fig:edx}b).\\  
To investigate the nontrivial diffusion properties of KTP in more detail, we fabricated a new set of waveguides for 5\,min at 330$^\circ$C. The results of the comparison are shown in Fig. \ref{fig:330_5_800}a). Two main features can be recognized. Firstly, the diffusion depth in waveguides larger than  3.0\,\textmu m is not influenced by the diffusion time. 
This is in agreement with the result reported by Bierlein et al. \cite{bierlein2}. Secondly, for short exchange times, the waveguide depths have a lower variability and seem uncorrelated with the waveguide widths. 
For getting a deeper understanding we also had a look at the surface concentration of the waveguides. In a simple model we assume an infinite reservoir for the exchanged waveguides and would expect the surface concentration to be independent of the exchange time. This would indicate that the surface concentration for 5\,min and 60\,min should be the same, but this is not the case. 
Fig. \ref{fig:330_5_800}b) shows the ratio of exchanged potassium ions with rubidium ions $\frac{c_{Rb_0}}{c_{K_{mean}}}$ where $c_{Rb_0}$ is the surface concentration of rubidium at the beginning of the waveguide. $c_{K_{mean}}$ is the mean value of the concentration of potassium inside the KTP bulk material analysed with a second linescan outside the waveguide. Therefore a completely exchanged surface has a ratio of 1. The 5\,min exchanged waveguides have a mean ratio over all waveguide widths of 0.29$\pm$0.03, while the mean ratio of the 60\,min exchanged waveguides with a value of 0.66$\pm$0.06 is more than twice as high.
We can also see that the surface concentration is independent of the waveguide width, even though the waveguide depth changes with the width. We attribute this effect to an increased stress component in the transition area between KTP and rubidium exchanged KTP. Stress can be expected due to the large lattice mismatch between KTP and rubidium titanyl phosphate (RTP) \cite{zumsteg}. We expect that for small waveguides we will find more stress because these transitions areas are located closer together. This stress could influence the diffusion properties and therefore the diffusion depth.\\
Furthermore, we produced waveguides exchanged for 60\,min at 330$^\circ$C and for 60\,min at 350$^\circ$C to see the influence of the temperature. In principle we would expect deeper waveguides with higher temperatures and that is exactly what we see in Fig. \ref{fig:330_5_800}c). The waveguides produced at 350$^\circ$C are deeper than the ones produces at 330$^\circ$C. Moreover we can see that the trend of deeper waveguides with narrower widths is also present for higher temperatures.\\
EDX is only sensitive to the atomic stoichiometry, while it cannot provide any information on stress and other structural properties, e.g. disorder or defects. Raman spectroscopy can be used to search for exchange-induced stress as it was previously demonstrated \cite{Pea-Rodrguez2012,Tejerina2014,DeWolf1996,Wolf1999,Panfilova2009}.	\\
	
\begin{figure}
		\centering
		\includegraphics[width=12cm]{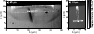}
		\caption{Raman image of a 40 \textmu m (a) and 1.5 \textmu m (b) wide waveguide. The images have been generated based on the spectral intensity in the 140 to 170 cm$^{-1}$ range. On the 40 \textmu m waveguide only the waveguide sides show an increase in signal, while the central regions show almost the same intensity as the surrounding bulk material.}
		\label{fig:Raman} 
	\end{figure}

	\paragraph{Raman spectroscopy} $~~$\\
To investigate the waveguide structures for stress, Raman imaging has been performed on a 40\,\textmu m and 1.5\,\textmu m wide waveguide. 
Fig. \ref{fig:Raman} shows the plots of the integrated intensity in the spectral region between 140 and 170 cm$^{-1}$ on the b-c faces of the waveguides similar to the experimental geometry on the EDX measurement. This spectral region is mainly composed of Raman bands associated with potassium or rubidium ions, respectively \cite{Kugel1988,Watson1991,Rusing2016a}. While the peak positions in this broad spectral region will be affected by the ion exchange, the integrated intensity is expected to be independent of the Rb concentration. However, stress can influence the intensity of Raman bands \cite{Fontana2008}. We find that the Raman images are completely different in the two waveguides analyzed: in the smaller waveguide, the intensity enhancement covers completely the exchanged region, while in the wider one it is localized only at the waveguide/bulk interface. Moreover, in 40\,\textmu m wide waveguide, the signal extends deep in the sample along the waveguide edges and becomes much shallower as it approaches the center of the waveguide. It is reasonable to assume that the stress is mostly localized at the interface between the exchanged waveguide and the pure bulk crystal. The reason is that the anisotropical diffusion will dramatically increase the unit cell dimensions in the waveguide. From the measurement of the 40\,\textmu m wide waveguide, we can see that the Raman signal has almost vanished over an up to 3\,\textmu m away from the waveguide/bulk interface. This can be interpreted as a gradual relaxation of stress in the crystalline structure. These observations suggest that all the waveguides in the range 1.5-4.5\,\textmu m are probably affected by the stress and thus the diffusion dynamic inside the waveguide is poorly modeled by Eq (1).

	\section{Simulations}
Seeing that the measured stress from the Raman spectroscopy shows unique behaviour for narrow and broad waveguides, we investigate if a stress-dependent diffusion coefficient can explain the data reported in Fig. \ref{fig:330_5_800}.
A reasonable assumption is to connect the diffusion coefficient $D$ to the local concentration of Rb in the crystal, as a higher Rb content results in a greater distortion of the crystalline lattice. For simplicity, we assume an exponential dependence of the diffusion coefficient $D$ on the concentration $c$ of Rb 
	\begin{equation}
	D(c) = D_0 \cdot \textrm{exp}\left[k\cdot c\right].
	\label{eq:d_c}
	\end{equation}
The parameter $k$ is related to how much stress is present locally or, equivalently, how much the diffusion characteristics are affected by the local Rb concentration. Higher stress is associated to a higher value of $k$, such that the diffusion is more affected by the Rb concentration.
	
\begin{figure}[ht!]
		\centering
		\includegraphics[width=7cm]{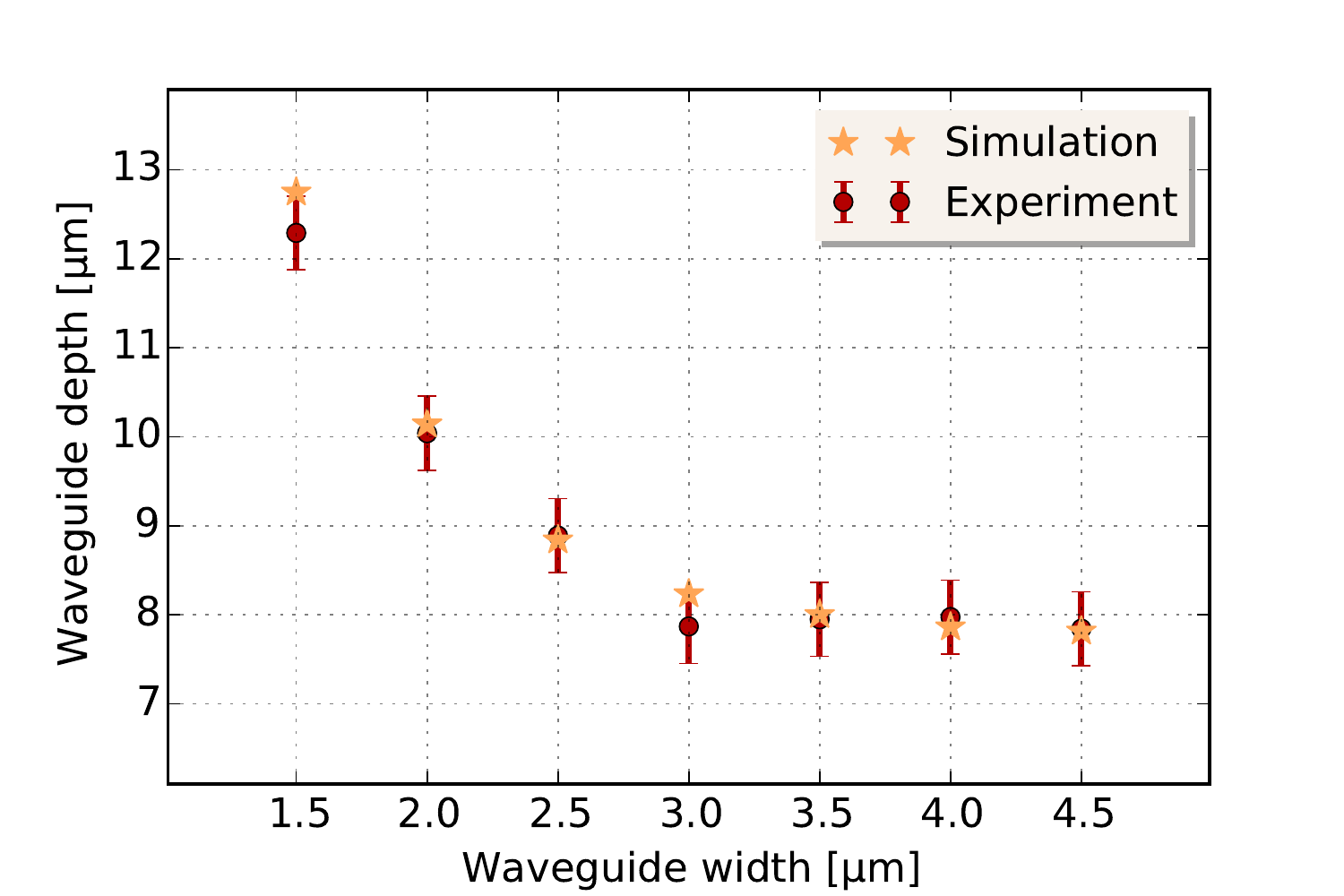}
		\caption{Simulations with stress-dependent diffusion coefficients displays the trend of deeper diffusion depth with narrower waveguide width.}
		\label{fig:simulation}
	\end{figure}

Equation \eqref{eq:d_c} assumes that the whole waveguide is characterized by the same level of stress $k$. However, the Raman measurements show that stress is localized nearby the edge of the waveguide, and is gradually relaxed over 2-3 $\mu$m. We can model this observation assuming a spatially varying $k$, which is maximum at the edges of the waveguide and decays towards $k=0$ in the centre of the waveguide. A possible functional that models this behaviour is 
	\begin{equation}
	k(y) = k_0\left[1+\frac{1}{2}\mathrm{erf}\left(\frac{y-\frac{w}{2}}{\sigma}\right)-\frac{1}{2}\mathrm{erf}\left(\frac{y+\frac{w}{2}}{\sigma}\right)\right],
	\label{eq:kappa}
	\end{equation}
where $y = 0$ corresponds to the centre of the waveguide and $w$ is the waveguide width. The parameter $\sigma$ determines how quickly stress is relaxed towards the centre of the waveguide, while $k_0$ determines the strength of the stress at the edges of the waveguide. \\
The model has three free parameters, $D_0$, $k_0$ and $\sigma$. We to set $D_0 = 0.25 \mu\text{m}^2/$min in order to match the waveguide depths for the wider waveguides, since their constant depths suggest that stress has little impact in these structures ($k=0$). From the Raman measurements, we set $\sigma = 1 \mu$m, as we have seen that the stress is relaxed over such distance.\\
We follow the method presented in \cite{crank_book} to solve the diffusion equation with a concentration-dependent $D$. Since the EDX data are taken at the centre of the waveguide, we calculate the diffusion profile by evaluating $k$ at $y = 0$ from Eq. \eqref{eq:kappa} and solving the diffusion equation with $D = D_0\exp{[k(0) c]}$.\\
We optimize the parameter $k_0$ to match the measured and the simulated depth of the waveguides, for different nominal widths. 
The comparison between simulations with $k_0$=3.1 and the measured data is shown in Fig. \ref{fig:simulation}.
	
A few comments are necessary regarding the presented model. The relationship between the Rb concentration, the stress in the waveguide and the diffusion coefficient provided in Eqs. \eqref{eq:d_c} and \eqref{eq:kappa} have been chosen to be physically reasonable, but their real interplay is still unknown.
Moreover, this model cannot reproduce the measurements for the 5\,min-diffused waveguides.
In fact, it can be shown that, if the $D$ depends only on the Rb concentration, it is impossible to have identical diffusion depths for a 5\,min and a 60\,min diffusion \cite{crank_book}.
This suggests that the overall diffusion process in our waveguide structures changes over time. We speculate that the diffusion process consists of at least two successive steps. In the first step, which lasts approximately for the first 5min of the exchange, the diffusion is fast and stress is not critical in determining the concentration profile. In the second step, the overall diffusion slows down but, at the same time, stress starts becoming more relevant, thus determining a width-dependent diffusion depth. \\
Further measurements are needed to validate our speculations and improve the modelling. It will be important to characterize the waveguide depths and the Raman images of waveguides fabricated with diffusion times ranging from a few minutes to a few hours. This will help understanding whether the build-up of the stress is a long term dynamics or if it arises already in the first minutes of the diffusion, and it will help to map the time evolution of the diffusion coefficient. 
Finally, it is necessary to characterise the concentration profile at different positions within the waveguide, to monitor how much the presence of the stress nearby the waveguide edge affects the diffusion. Unfortunately, EDX is not particularly suited for this purpose, since the minimum dimensions of the interaction volume prevent the resolution of features smaller than $\sim$1.4\,$\mu$m.
Only with these additional measurement it will be possible to improve our understanding of the highly nonlinear diffusion of Rb into KTP.

\section{Conclusion}
In conclusion, we have characterized for the first time methodically the diffusion process for the realization of rubidium exchanged waveguides in KTP. We were able to show that the penetration depth depends on the width of the waveguide. We find that the narrower we produce our waveguides the deeper they become in comparision to a simple one dimensional diffusion model. Moreover we see that the variation of the penetration depth with the width is stronger for higher temperatures and longer diffusion times. 
Our preliminary Raman analysis reveals the presence of stress, in particular at the edge of the waveguides. This is likely to modify the diffusion dynamics of Rb and could explain the breakdown of the simple 1D diffusion model. The underlying mechanisms, however, are not fully understood yet and further analysis, e.g. Raman spectroscopy or SHG microscopy, is necessary.\\
This work shows complex rubidium exchange behaviour needs further investigations for a complete explanation of the empirical observations. But the first Raman measurements suggest that stress in the waveguide might be a critical factor in the diffusion process. Further understanding of these effects will be crucial for the realization of improved waveguide performance in this material. A more complete understanding of the rubidium exchange process will enable the fabrication of more homogeneous and uniform KTP waveguides.

\section*{Funding}
Deutsche Forschungsgemeinschaft (SFB - TRR 142, Projektnummer 231447078).

\section*{Acknowledgments}
Funded by the Deutsche Forschungsgemeinschaft (DFG, German Research Foundation)	- Projektnummer 231447078 - TRR 142.

\section*{Disclosures}
The authors declare no conflicts of interest. 	

\bibliography{sample}
\end{document}